\documentclass[conference]{IEEEtran}

\usepackage{graphicx} 
\usepackage{subfigure}
\usepackage{picins}
\usepackage{fancyhdr}

\IEEEoverridecommandlockouts
\IEEEpubid{\makebox[\columnwidth]{978-1-5090-4130-5/16/\$31.00~
\copyright2016
IEEE \hfill} \hspace{\columnsep}\makebox[\columnwidth]{ }}

\newcommand{\Fig}[1]{Fig.~\ref{#1}}

\title{We Hear Your Activities through Wi-Fi Signals}

\author{Fang-Jing~Wu and G\"urkan~Solmaz \\
NEC Laboratories Europe, Heidelberg, Germany\\
fang-jing.wu@neclab.eu, gurkan.solmaz@neclab.eu
}

\fancypagestyle{IEEE_Green_open_access_footer}{%
  \fancyhf{}
  
  \fancyfoot[C]{\footnotesize \copyright2016 IEEE. Personal use of this material is permitted. Permission from IEEE must be obtained for all other uses, in any current or future media, including reprinting/republishing this material for advertising or promotional purposes, creating new collective works, for resale or redistribution to servers or lists, or reuse of any copyrighted component of this work in other works. DOI: 10.1109/WF-IoT.2016.7845478}
}

\begin{document}

\maketitle
\thispagestyle{IEEE_Green_open_access_footer}

\begin{abstract}
In this paper we focus on the problem of human activity recognition without identification of the individuals in a scene. We consider using Wi-Fi signals to detect certain human mobility behaviors such as stationary, walking, or running. The main objective is to successfully detect these behaviors for the individuals and based on that enable detection of the crowd's overall mobility behavior. We propose a method which infers mobility behaviors in two stages: from Wi-Fi signals to trajectories and from trajectories to the mobility behaviors. We evaluate the applicability of the proposed approach using the StudentLife dataset which contains Wi-Fi, GPS, and accelerometer measurements collected from smartphones of 49 students within a three-month period. The experimental results indicate that there is high correlation between stability of Wi-Fi signals and mobility activity. This unique characteristic provides sufficient evidences to extend the proposed idea to mobility analytics of groups of people in the future.
\end{abstract}


\section{Introduction}
Human activity recognition is a research challenge mainly tackled by computer vision researchers using videos captured by security cameras to understand individuals' behaviors. However, recording people's activities with their identities brings certain privacy concerns due to lack of anonymity. Moreover, in certain scenarios, the cameras may not be deployed due to governmental restrictions or other concerns. Another aspect of this problem is that the processing of frames of videos, tracking each individual's movement, and understanding their behavior is only possible with considerable computational and communication overheads. On the other hand, Wi-Fi signals can be captured by Wi-Fi sniffers or by deploying Wi-Fi access points (APs) in an area. Furthermore, smartphones with Received Signal Strength Indicators (RSSIs) can be used for mobile sensing of the Wi-Fi APs and deciding the current behaviors of the smartphone users. Therefore, we believe that detection of people's mobility behaviors through Wi-Fi signals is necessary and relevant alternative solution for the human activity recognition.

Among different types of human activities, this work focuses on the mobility activities particularly for \emph{pedestrian mobility activities} without using motor vehicles (e.g., stationary, walking, running activities). Transportation by buses or high-speed trains in larger areas are not considered in the scope of this paper. Some of existing efforts focus on pedestrian mobility~\cite{Solmaz15} in theme parks for enhancing public safety. We aim at pedestrian movements to recognize stationary, walking, and running activities in small-scale areas such as campuses or shopping malls. Due to the popular use of connected mobile devices such as smartphones and tablet computers, human mobility behaviors can be captured through Wi-Fi signals. Understanding movement patterns of people is important in many areas such as traffic engineering, crowd management, disaster management, urban planning, epidemic modeling, and mobile networks. For instance, considering a natural or man-made disaster in an area, finding effected people who cannot move, need emergent help or stuck in pedestrian or vehicle traffic is of utmost importance for successful save, rescue and evacuation operations.

This work proposes a mobility activity inference model based on Wi-Fi signals to detect three mobility activities: stationary, walking, and running. Since multi-modal sensing approaches may suffer from the missing data problem due to unavailability of data sources or low sampling rates of sensors, we design a two-phase algorithm to infer mobility activity. Specifically, we exploit the correlation between stability of Wi-Fi signals and moving speeds to compensate for the lack of data in the first phase and then infer mobility activities using classifiers in the second phase. The major contribution of this paper is to figure out the correlation of Wi-Fi signals and human mobility activities which can be applied to a more flexible system without the Global Positioning System (GPS) localization technologies and other types of data sources.

The remainder of this paper is organized as follows. The second section reviews the existing research on human activity recognition. The third section explains our proposed models for mobility activity inference. The experiments are presented in the fourth section. The last section concludes this paper.

\section{Related Work} \label{Sec:Related_Work}
Recently various research studies have been conducted related to human activity recognition. Some of these studies are focused on using body-worn sensors~\cite{Bulling14,Lara13} or smartphones~\cite{Incel13} while most literature can be found in the computer vision field~\cite{Aggarwal14,Ke13}. Bugdol et al.~\cite{Bugdol16} propose the use of smartphone sensors (e.g., accelerometers, microphones) for recognizing human activities such as riding bike, driving car, sitting, and walking. They propose Support Vector Machine (SVM) for classification of the inputs from various sensors. Bayat et al.~\cite{Bayat14} propose using built-in triaxial accelerometers of smartphones and they test their activity recognition model for people performing physical activities such as slow walking, running, fast walking, climbing stairs, and aerobic dancing. Wu et al.~\cite{Wu2014_UrbanMobilitySense} use built-in sensors on smartphones to collect GPS, Wi-Fi, GSM cells, and accelerometers for transportation mode inference. An unsupervised learning approach by hierarchical clustering and density-based spatial clustering of applications with noise (DBSCAN) is proposed by Kwon et al.~\cite{Kwon14} using smartphone sensor data. Instead of using smartphone data, Wi-Fi signals are used by Wang et al.~\cite{Wang15} for recognition of activities such as walking, running, brushing teeth, falling, opening refrigerator, and so on. This approach is an alternative for understanding body movements such that it does not require wearing body sensors. Their Hidden Markov Model (HMM)-based model provides accurate understanding of the human body movement recognition. Alvarez et al.~\cite{Alvarez13} tackle the similar problem of body posture recognition with Wi-Fi signals and accelerations.

Our study differs from the aforementioned ones in the sense that we aim to understand the micro movement behaviors of a person, while others focus on body movements or large-scale movements made by motor vehicles. We aim to use Wi-Fi signals which always exist in surroundings with additional resources (if available) such as smartphone sensors (e.g., GPS, Bluetooth, and accelerometers) for inferring mobility activities. Moreover, our method is more flexible when some of the data sources are not available.

\section{Mobility Activity Inference Model}\label{Sec:System Model}

\subsection{Human Mobility}

Understanding human mobility brings great potential benefits for networked systems, epidemics, traffic control, urban planning, and disaster management. Scientific research on human mobility gained momentum with new findings for better understanding of human mobility~\cite{Brockmann06,Gonzalez08}. With the help of tracking thousands of cell phones, researchers are able to find the limits of predictability~\cite{Song10} and main properties of human movements such as the walking distance distributions and diffusion. This leads to promising efforts on developing new mobility models which are able to represent the human movement patterns more realistically. Although people can be tracked with videos and built-in smartphone technologies such as GPS, this does not seem feasible for various environments due to privacy concerns. Moreover, it is very hard to gather GPS data in some environments in defined period of times such as the movement of people during disasters in a particular city. Another aspect of human mobility is the fact that the movement of people depend on their environment. For instance, a person who lives in a metropolitan area goes to work everyday and come back home with vehicles, while in another environment such as in a campus environment the person travels by walk. Therefore, scenario-specific consideration is necessary for having accurate representations of human mobility patterns. Furthermore, pedestrian mobility and the mobility of vehicles are made by very different means such that the maximal moving speed of pedestrians is much slower than the speed of people on motor vehicles. Thus, this distinction on the pedestrian mobility is necessary.

While there exist some studies related to human mobility, realistic understanding of human mobility stays a major challenge for various scenarios ranging from human mobility in continental scale to mobility inside buildings. Especially, with the advancements in mobile computing and upcoming technologies such as 5G and the Internet of Things (IoT), human mobility will have much more importance than it nowadays has. We focus on the human mobility in the urban environments such as city squares, stadiums, airports, and theme parks as well as the natural and man-made disaster scenarios. To understand the movement of the crowd in pedestrian flows or in city traffic and finding the people who need assistance, certain human mobility behaviors can be inferred with the help of Wi-Fi signals. For instance, by understanding the people who mostly stay in the same place, we can find the people in need of a help during disasters.

Our first goal is to infer 3 simple human mobility activities using Wi-Fi signals: stationary, walking, and running. We define the \emph{stationary} mode as the mode of a person who stays in a certain radius for a certain period of time. These radius and time values can be set as parameters and they may differ according to the environment (e.g., city square or airport) and the conditions (e.g., ordinary or emergency situations). \emph{Walking} is the case when the person seems to move but the Wi-Fi signals do not fluctuate so much. Our logic is to map these Wi-Fi signal fluctuations to the actual distance changes so that one can understand that the person is in the walking mode. There is a certain estimated speed parameter such that if the estimated movement passes a certain speed in a defined period of time, the person is considered in the \emph{running} mode.

\begin{figure}
\centering
\includegraphics[width=\columnwidth]{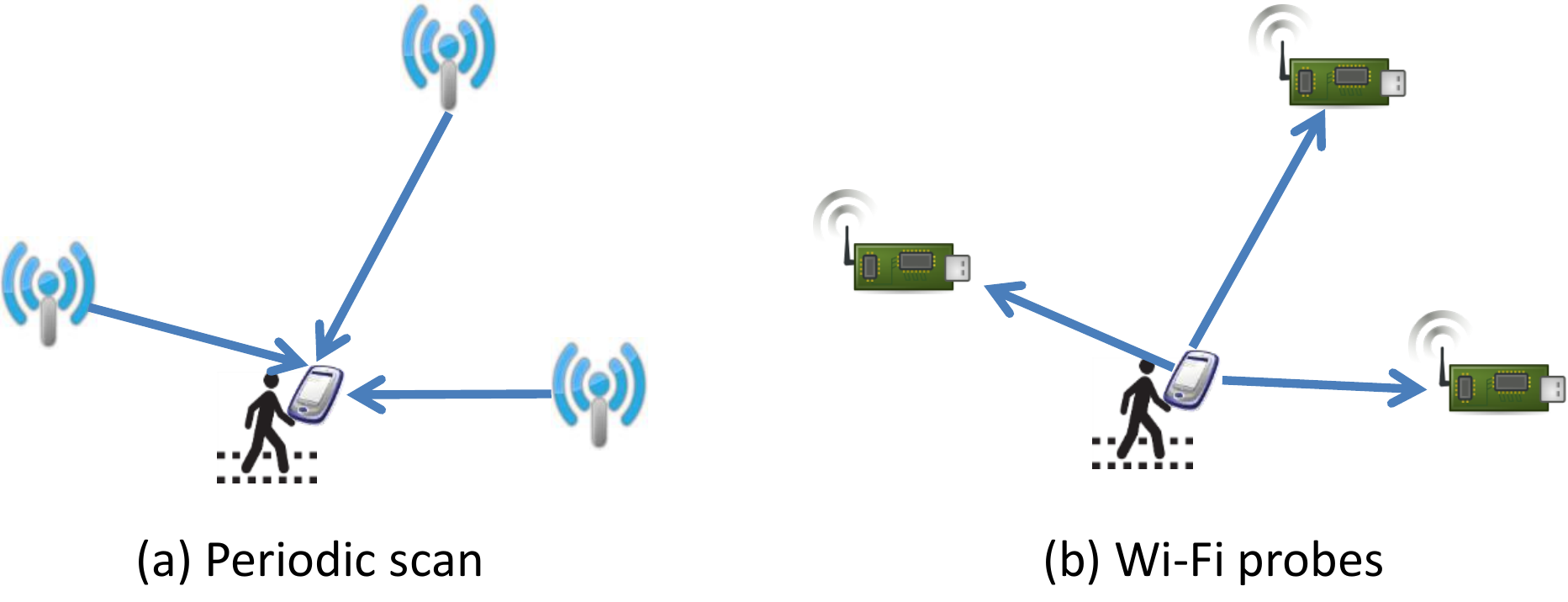}
\caption{Two data-collection mechanisms.} \label{Fig:data-collection-mechanism}
\end{figure}

\subsection{Data Collection Mechanisms}
This work investigates the correlation between RF signal fluctuation and human mobility activities including stationary, walking, and running. \Fig{Fig:data-collection-mechanism} shows two data-collection mechanisms for capturing RSSIs between Wi-Fi signal transmitters and receivers. \Fig{Fig:data-collection-mechanism}(a) shows that a mobile application running on the smartphone collects RSSIs from ambient visible Wi-Fi APs periodically. In contrast to active scan, \Fig{Fig:data-collection-mechanism}(b) uses multiple Wi-Fi sniffers to collect Wi-Fi probes broadcasted by smartphones from time to time. There are different essential limitations for two data-collection mechanisms. The first one relies on mobile users' will to install the mobile application in their smartphones, while the latter one relies on additional sensor deployments. This work first analyzes the data collected by active scan, and we will take RSSIs from Wi-Fi probes into account in our future work.

\subsection{Activity Classification Models}

\begin{figure}[!t]
\centering
\subfigure[GPS speed.]{\label{Fig:GPS-speed-user3}
\includegraphics[width=\columnwidth]{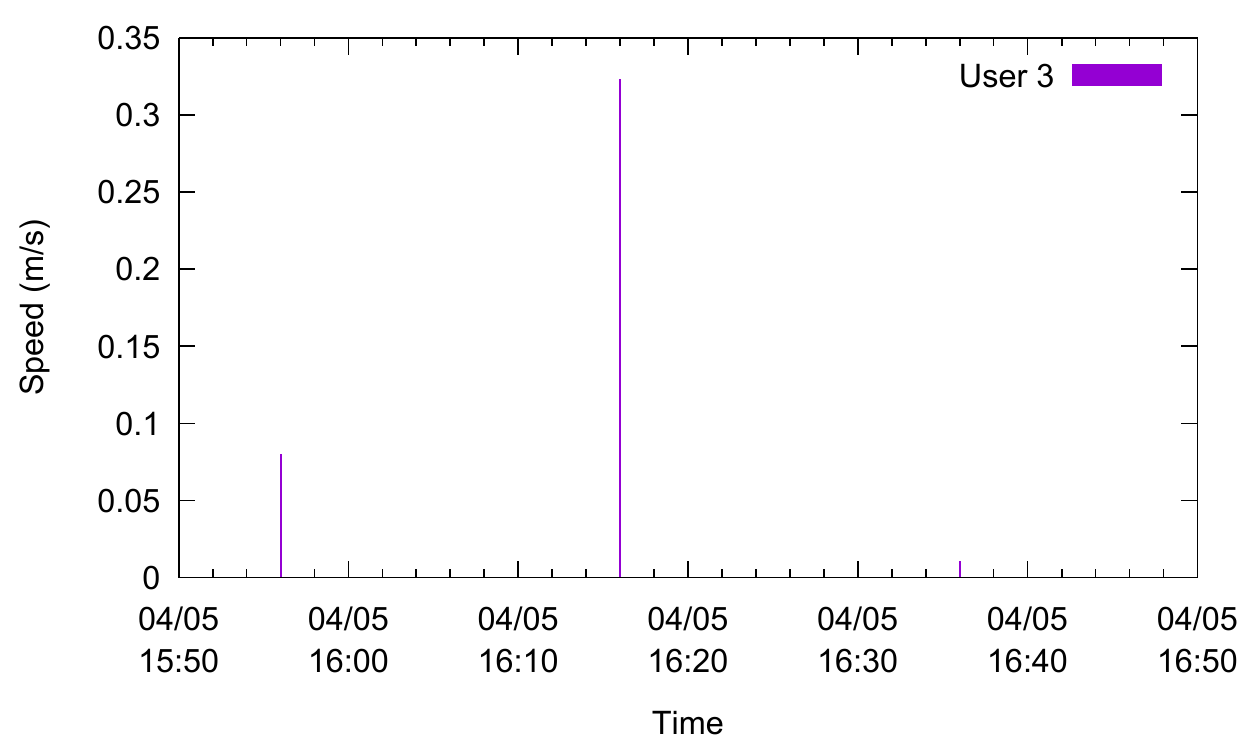}}
\subfigure[Stability of Wi-Fi signals.]{\label{Fig:stability-user3}
\includegraphics[width=\columnwidth]{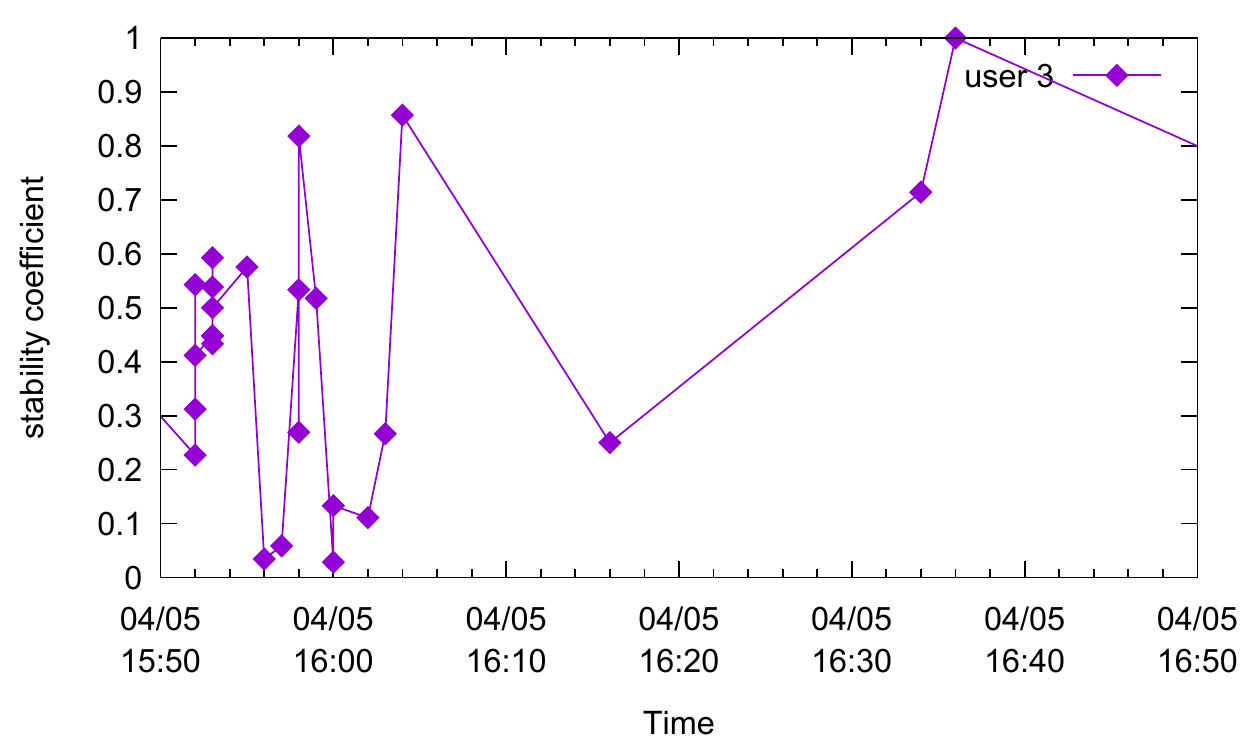}}
\hspace{0.01in}
\subfigure[Labelled activities by a mobile app for mobility activity detection.]{\label{Fig:fig_activity3}
\includegraphics[width=\columnwidth]{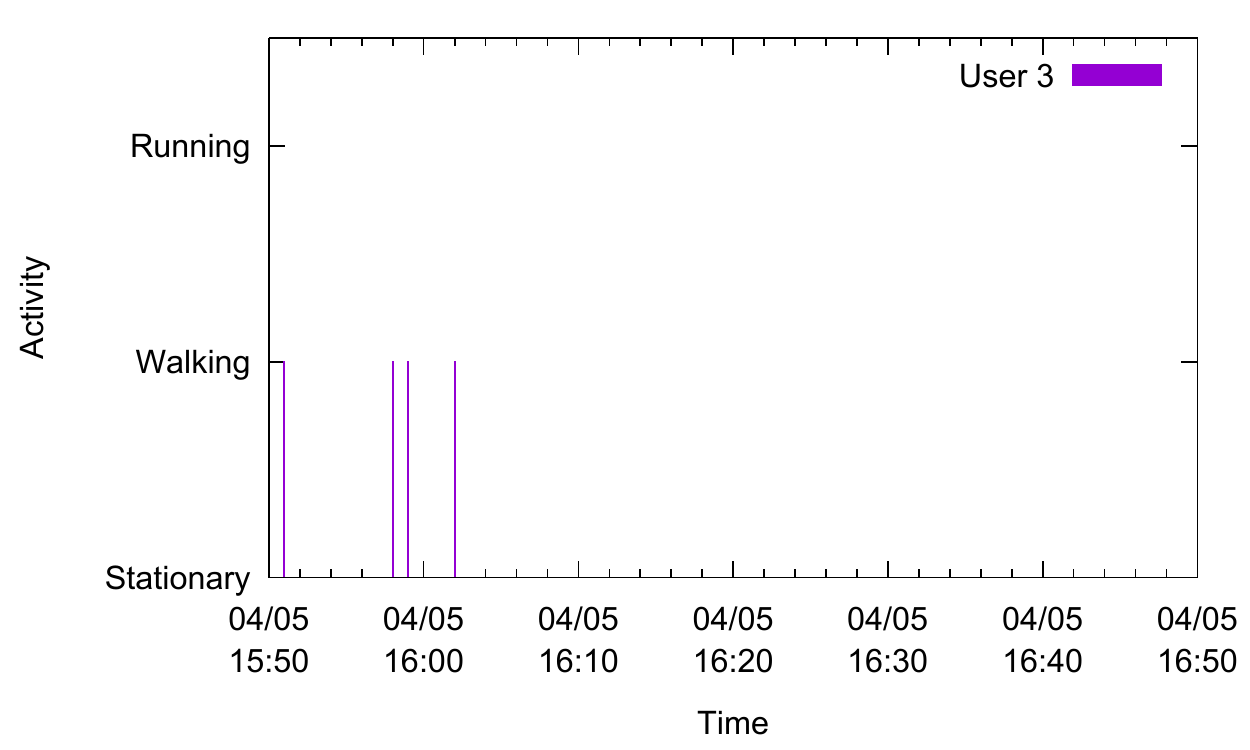}}
\hspace{0.01in}
\caption{Observations on the correlation between stability of Wi-Fi signals, moving speeds, and labelled activities.} \label{Fig:observations}
\end{figure}

Since some types of sensing data may suffer from the missing data problem due to low sampling rates or unavailability of data sources, we propose a two-phase classification model to inference mobility activities where the first-phase is to solve the missing data problem and the second-phase performs mobility activity inference using with sufficient information. For example, when GPS does not work in indoor places or Wi-Fi sniffer is considered for data collection, moving speeds are missing most of time. Therefore, we make observations on GPS moving speeds, stability of Wi-Fi signals, and labelled activities based on the StudentLife dataset~\cite{Wang14} to understand the correlation between the multiple types of data modalities, as shown in \Fig{Fig:observations}. To save energy of smartphones, GPS works at an extremely low sample rate of 1/600 Hz. Thus, the missing GPS data problem becomes serious. As it can be seen that the stability coefficients are low when either the labelled activities indicate walking or GPS moving speeds have significant changes. We can see that the stability of Wi-Fi signals can provide sufficient clues to infer mobility activities. Here, we consider the first data-collection mechanism to compute the stability coefficient of Wi-Fi signals between two consecutive sets of ambient Wi-Fi signals detected by the smartphone denoted by $X$ and $Y$ using Jaccard index as follws
\[
S(X, Y)= \frac{|X\bigcap Y|}{|X \bigcup Y|},
\]
where the stability coefficient is between 0 and 1. A lower stability coefficient means that Wi-Fi signals fluctuate significantly. In contrast, a higher  stability coefficient means that Wi-Fi signals are more stable.

\begin{figure}
\centering
\includegraphics[width=0.85\columnwidth]{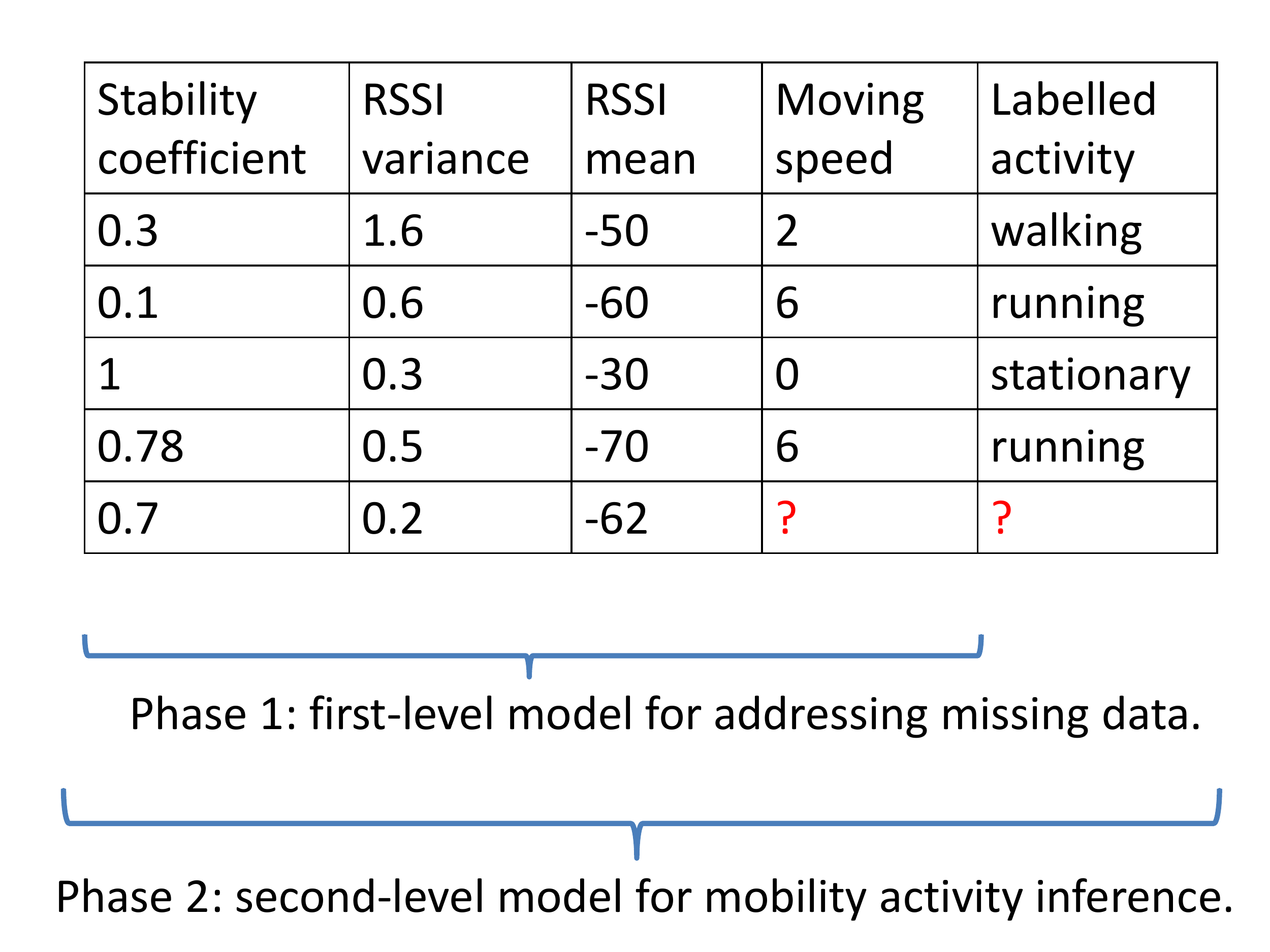}
\caption{Two-stage activity inference.} \label{Fig:two-stage-inference}
\end{figure}

Based on the above observations, \Fig{Fig:two-stage-inference} illustrates the idea of the proposed two-stage activity inference. Since Wi-Fi data is always available in the surrounding, we use Wi-Fi fluctuating information to learn the first-level model for inference of moving speeds. Here, we consider the Gaussian process regression as the first-level model to address the missing data problem, where complete data including Wi-Fi fluctuations and moving speeds is considered as the training dataset for a regression function. The moving speeds in incomplete dataset will be estimated through the regression function. Then, both Wi-Fi information and moving speeds are taken into account to learn the second-level classification model for human mobility inference. We consider three different classifiers, J48Tree, Naive Bayes, and Random Forest, as the the second-level classification model to compare the accuracy. Note that the second-level classification model is not limited to the three classifiers for a more flexible approach.

\section{Experimental Results}\label{Sec:Experiments}
\subsection{Dataset Description}

We analyze Wi-Fi scan traces as well as GPS traces from 49 students in the StudentLife dataset~\cite{Wang14}. The dataset contains measurements from smartphones of students in Dartmouth College including various sensor data such as audio, phone calls, phone charge. Measurements of Wi-Fi signals, GPS traces, and the activity inference datasets are used in this experimental study. The speed of each person is computed over all the GPS trajectories to be able to analyze their correlations with the Wi-Fi signal strengths.

The dataset is filtered to exclude the inaccurate measurements. For instance, the location data based on connected cell towers in the location data files is removed. For the activity inference, the \emph{unknown} parts are excluded from the dataset. Description of the data set in terms of data sizes (number of entries) for each dataset and other related information can be seen in Table~\ref{tab:DatasetDescription}. Minimum and maximum data sizes mean the number of entries for the users with smallest or largest data among all 49 users. The numbers included in the table correspond to the modified dataset after filtering.

\begin{table}
   \centering
    \caption{Dataset description.}
        \begin{tabular} {|p{5.0cm}|p{2.3cm}|} \hline
            Number of users & 49 \\ \hline
            Accelerometer sampling time ($\approx$)  & 2-3 sec \\ \hline
            Measurement start date & March 27, 2013 \\ \hline
            Measurement end date & June 5, 2013 \\ \hline
            Movement data provider & Wi-Fi, GPS \\ \hline
            Min activity inference data size & 163113 \\ \hline
            Max activity inference data size & 1048576 \\ \hline
            Min GPS data size & 889 \\ \hline
            Max GPS data size & 6536 \\ \hline
            Min Wi-Fi data size & 10350 \\ \hline
            Max Wi-Fi data size & 373062 \\ \hline
                \end{tabular}
        \label{tab:DatasetDescription}
\end{table}

\subsection{Cross-modal Data Analyses}

\begin{figure}
\centering
\subfigure[GPS speed.]{\label{Fig:GPS-speed}
\includegraphics[width=0.9\columnwidth]{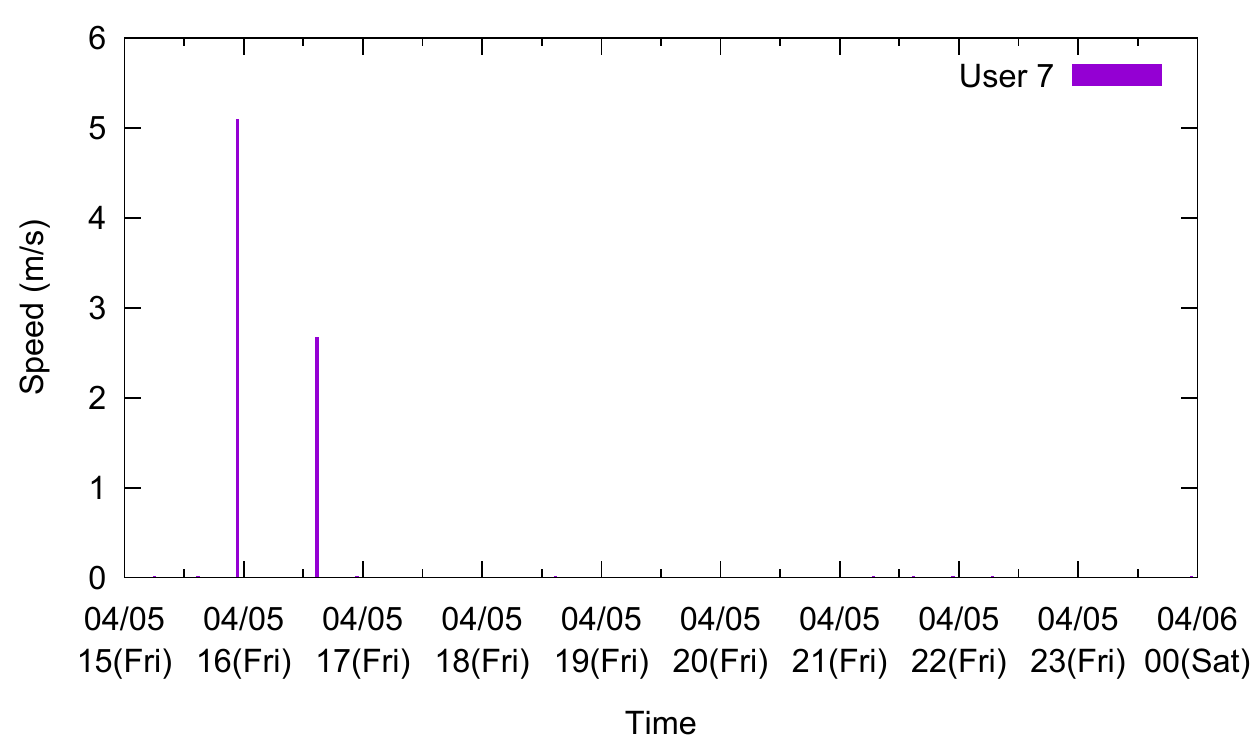}}
\hspace{0.01in}
\subfigure[activity.]{\label{Fig:activity}
\includegraphics[width=0.9\columnwidth]{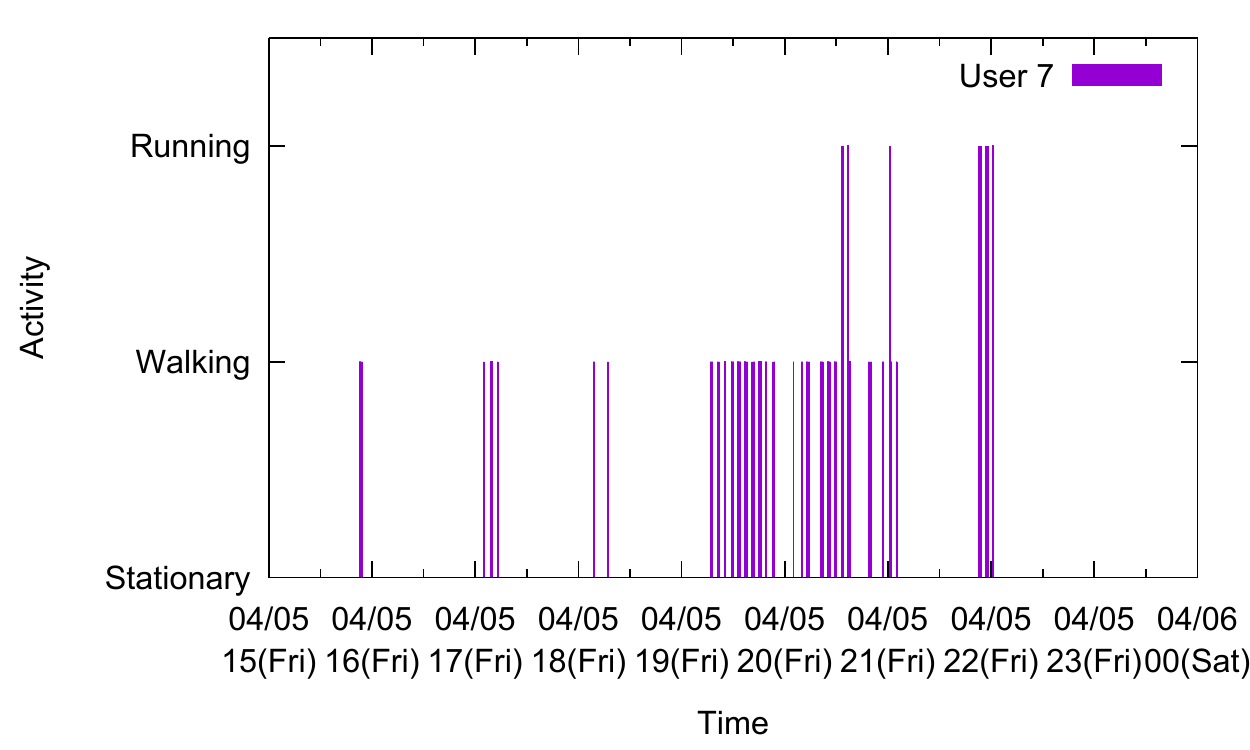}}
\subfigure[Number of ambient Wi-Fis.]{\label{Fig:wifi_scan_results}
\includegraphics[width=0.9\columnwidth]{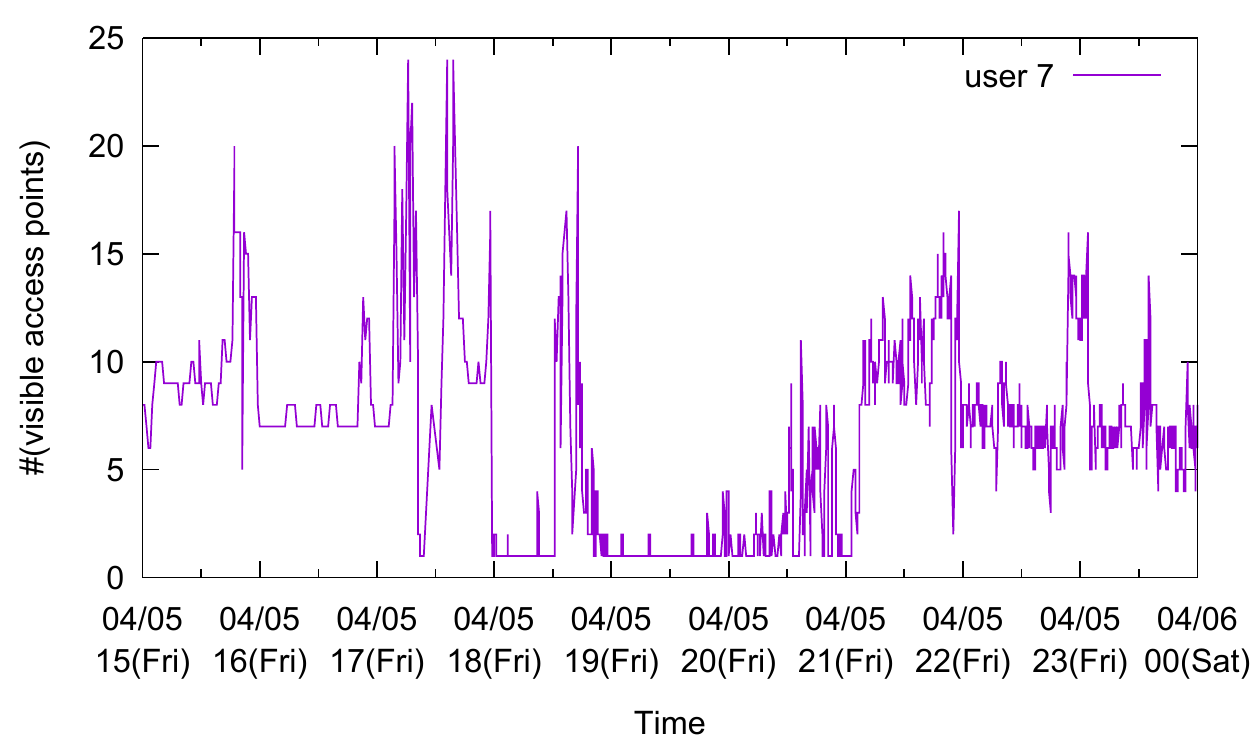}}
\subfigure[RSSI.]{\label{Fig:RSSIs}
\includegraphics[width=0.9\columnwidth]{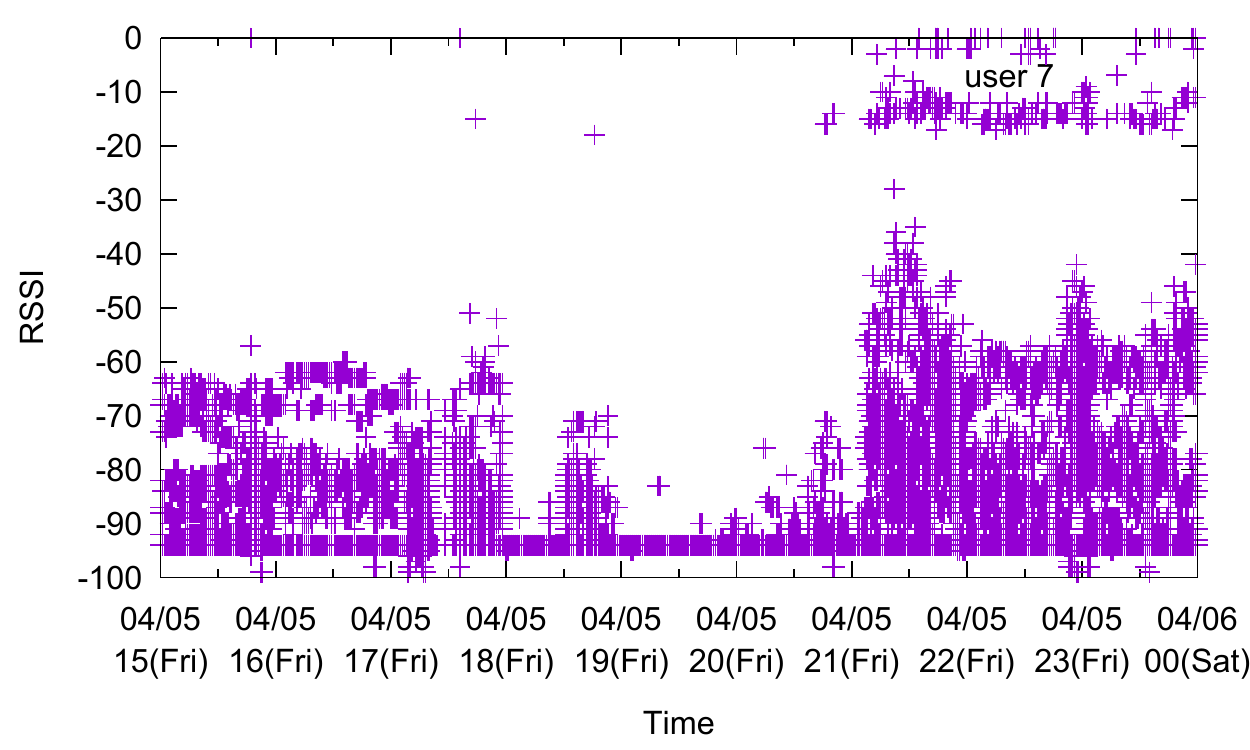}}
\caption{Experimental results.} \label{Fig:Experimental-results}
\end{figure}

First, we discuss the correlation and feature patterns of cross-modal data including GPS data, labelled activities, number of ambient visible Wi-Fi APs, and distribution of RSSIs. The data analysis results are presented in \Fig{Fig:Experimental-results}. As we can see in \Fig{Fig:Experimental-results}(a), GPS measurements are missing most of time because the user's smartphone probably cannot get GPS fixes. GPS data provides accurate information to detect large-scale mobility. However, for small-scale movements or micro mobility behaviour, such as walking or being stationary, GPS data is not sufficient. \Fig{Fig:Experimental-results}(b) shows labelled activities through an activity inference algorithm on the smartphone based on the accelerometer data. When the user changes mobility activity, the number of visible APs changes significantly as shown in \Fig{Fig:Experimental-results}(c). Similarly, the distribution of RSSIs changes when human mobility activity changes. The data analysis results provide hints to use Wi-Fi signal fluctuating information to compensate the missed data for mobility activity inference. For a more flexible data collection mechanism with multiple Wi-Fi sniffers, we can apply the proposed idea to mobility activity analytics of groups of people.

\subsection{Experimental Results of Activity Inference}
\begin{figure}
\centering
\includegraphics[width=\columnwidth]{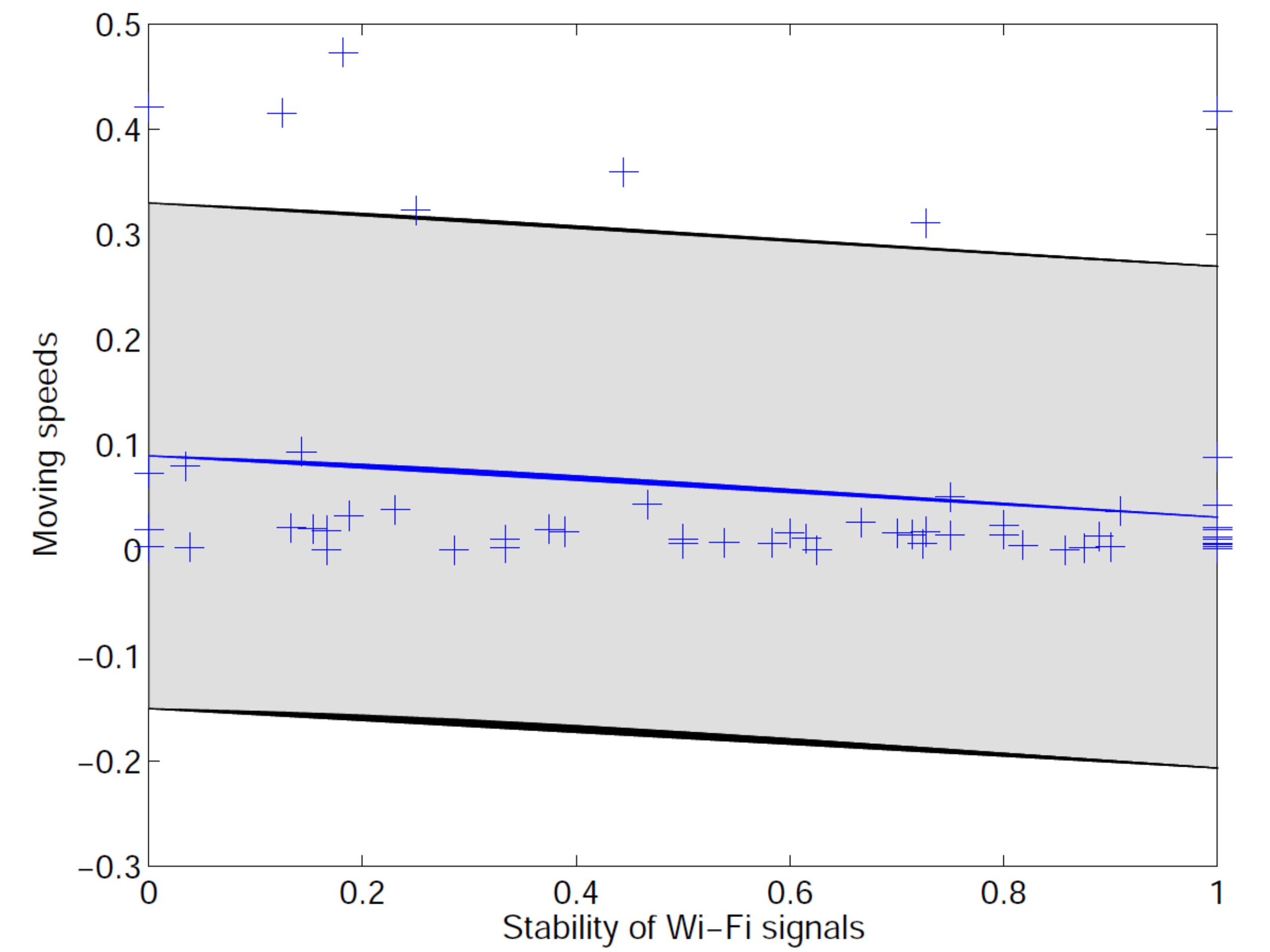}
\caption{Experimental results of the first-phase Gaussian process regression model for a particular user's data.} \label{Fig:GPR results}
\end{figure}

Finally, we discuss the experimental results of our first-level Gaussian process regression and the second-level classification models. \Fig{Fig:GPR results} shows the Gaussian process regression function between stability of Wi-Fi signals and moving speeds. As it can be seen, a lower value of stability coefficient indicates a higher moving speed. Since sensors and Wi-Fi antennas on different smartphones have different sensitivities and favourite visit places of users are different, stability coefficients resulted from different smartphones are probably different. There is no one-size-fit-all model when data collection is conducted by each individual's smartphone. In contrast, when Wi-Fi sniffers are used for collecting data, the sensitivity of these Wi-Fi antennas for collecting multiple users's Wi-Fi signals from mobile devices remains consistent.

\begin{figure}
\centering
\includegraphics[width=\columnwidth]{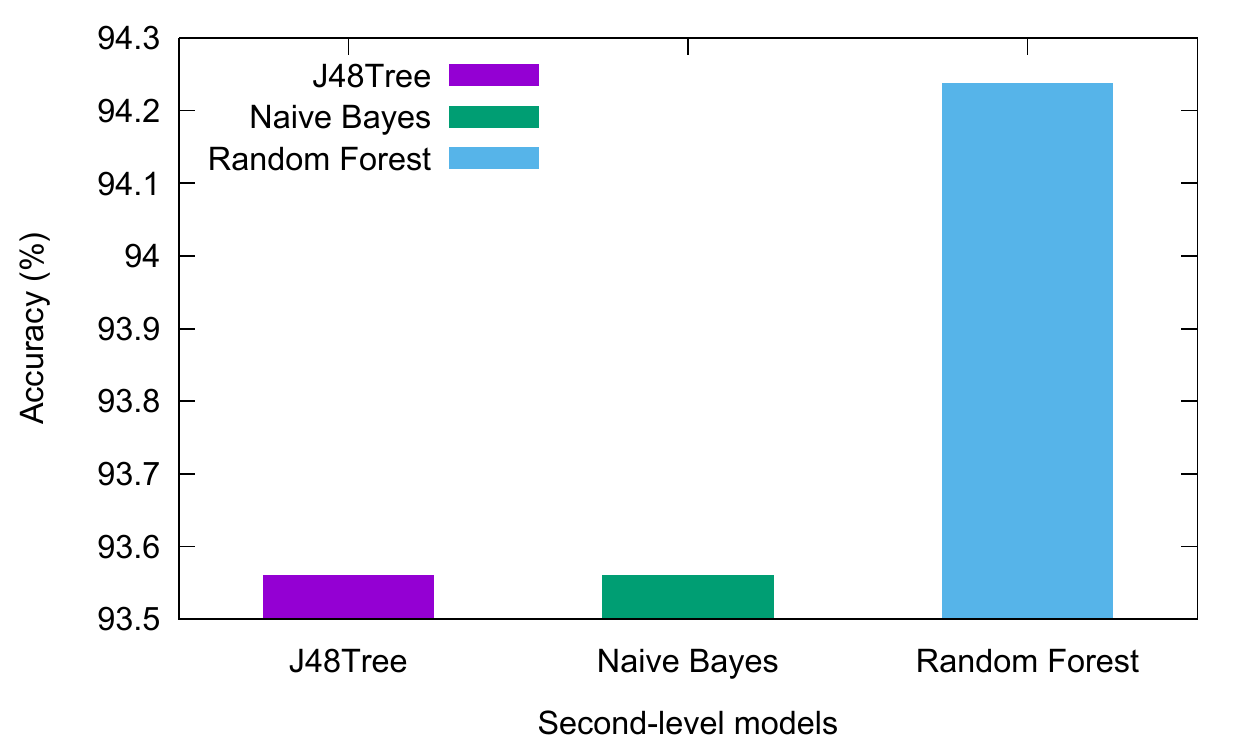}
\caption{Experimental results of our two-phase algorithm, where three classification models are considered in the second phase.} \label{Fig:ClassificationResults}
\end{figure}

Finally, we use Weka~\cite{Weka} to study the performance of the proposed mobility activity inference algorithm when different classifiers are applied to the second-level model. We compare the accuracy of mobility activity inference in this experiment, where J48Tree, Naive Bayes, and Random Forest, are considered as the second-level classifiers. Here, we consider the three classifiers since all of these three classifiers work well for numerical and categorical data and incur lower computation overhead which can support real-time detection tasks in the future. In this experiment, we split the dataset into two sub-datasets, where 50\% data for training and 50\% data for testing. \Fig{Fig:ClassificationResults} shows the experimental results. As it can be seen, the accuracy resulted from the Random Forest classifier is slightly better than the others.

\section{Conclusion}\label{Sec:conclusion}
In this paper, we focus on human activity recognition without the use of security cameras or body-sensors. We propose using Wi-Fi signals to detect simple human mobility behaviors of stationary, walking, and running. We propose a method which consists of two stages of human mobility inference from Wi-Fi signals to the mobility behavior. We analyze the proposed approach by using Wi-Fi, GPS, and accelerometer measurements from smartphones of 49 students.

As a future work, we intend to have extensive experiments with one or more people using Wi-Fi sniffers and Wi-Fi APs as well as smartphone sensors and calibrate our human activity inference model. Moreover, our future goal is to extend the proposed model and enable inference of crowd mobility behaviors in urban environments such as city squares and airports. We believe that this model can enable efficient crowd management and search-and-rescue operations in the future.

\section{Acknowledgment}
\parpic{\includegraphics[width=0.23\linewidth,clip,keepaspectratio]{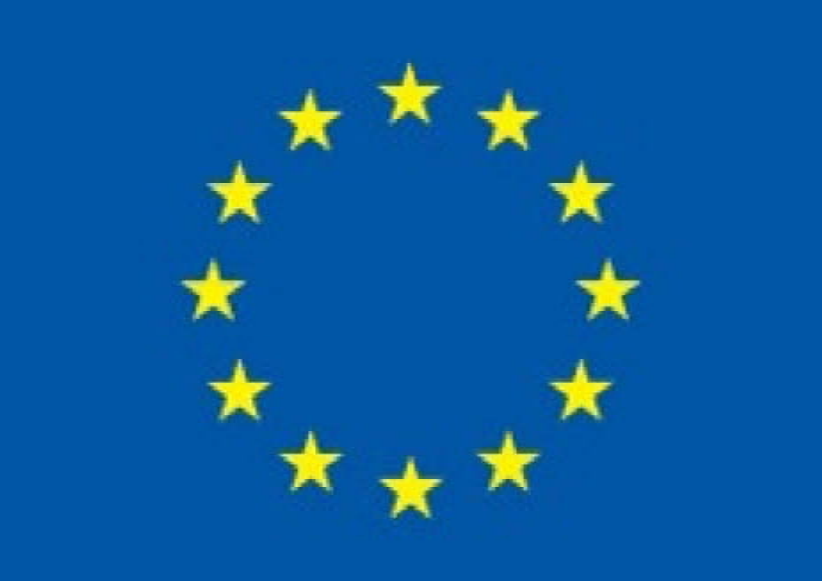}}
\noindent This work has received funding from the European Union's Horizon 2020 research and innovation programme within the project ``Worldwide Interoperability for SEmantics IoT" under grant agreement Number 723156.

\bibliographystyle{IEEEtran}

\begin{thebibliography}{10}
\providecommand{\url}[1]{#1}
\csname url@samestyle\endcsname
\providecommand{\newblock}{\relax}
\providecommand{\bibinfo}[2]{#2}
\providecommand{\BIBentrySTDinterwordspacing}{\spaceskip=0pt\relax}
\providecommand{\BIBentryALTinterwordstretchfactor}{4}
\providecommand{\BIBentryALTinterwordspacing}{\spaceskip=\fontdimen2\font plus
\BIBentryALTinterwordstretchfactor\fontdimen3\font minus
  \fontdimen4\font\relax}
\providecommand{\BIBforeignlanguage}[2]{{%
\expandafter\ifx\csname l@#1\endcsname\relax
\typeout{** WARNING: IEEEtran.bst: No hyphenation pattern has been}%
\typeout{** loaded for the language `#1'. Using the pattern for}%
\typeout{** the default language instead.}%
\else
\language=\csname l@#1\endcsname
\fi
#2}}
\providecommand{\BIBdecl}{\relax}
\BIBdecl

\bibitem{Solmaz15}
G.~Solmaz and D.~Turgut, ``Pedestrian mobility in theme park disasters,''
  \emph{Communications Magazine, IEEE}, vol.~53, no.~7, pp. 172--177, July
  2015.

\bibitem{Bulling14}
A.~Bulling, U.~Blanke, and B.~Schiele, ``A tutorial on human activity
  recognition using body-worn inertial sensors,'' \emph{ACM Computing Surveys
  (CSUR)}, vol.~46, no.~3, p.~33, January 2014.

\bibitem{Lara13}
O.~D. Lara and M.~A. Labrador, ``A survey on human activity recognition using
  wearable sensors,'' \emph{Communications Surveys \& Tutorials, IEEE},
  vol.~15, no.~3, pp. 1192--1209, 2013.

\bibitem{Incel13}
O.~D. Incel, M.~Kose, and C.~Ersoy, ``A review and taxonomy of activity
  recognition on mobile phones,'' \emph{BioNanoScience}, vol.~3, no.~2, pp.
  145--171, 2013.

\bibitem{Aggarwal14}
J.~K. Aggarwal and L.~Xia, ``Human activity recognition from 3d data: A
  review,'' \emph{Pattern Recognition Letters}, vol.~48, pp. 70--80, October
  2014.

\bibitem{Ke13}
S.-R. Ke, H.~L.~U. Thuc, Y.-J. Lee, J.-N. Hwang, J.-H. Yoo, and K.-H. Choi, ``A
  review on video-based human activity recognition,'' \emph{Computers}, vol.~2,
  no.~2, pp. 88--131, 2013.

\bibitem{Bugdol16}
M.~D. Bugdol, A.~W. Mitas, M.~Grzegorzek, R.~Meyer, and C.~Wilhelm, \emph{Human
  Activity Recognition Using Smartphone Sensors}.\hskip 1em plus 0.5em minus
  0.4em\relax Cham: Springer International Publishing, 2016, pp. 41--47.

\bibitem{Bayat14}
A.~Bayat, M.~Pomplun, and D.~A. Tran, ``A study on human activity recognition
  using accelerometer data from smartphones,'' \emph{Procedia Computer
  Science}, vol.~34, pp. 450--457, 2014.

\bibitem{Wu2014_UrbanMobilitySense}
F.-J. Wu and H.~B. Lim, ``{UrbanMobilitySense}: A user-centric participatory
  sensing system for transportation activity surveys,'' \emph{IEEE Sensors
  Journal}, vol.~14, no.~12, pp. 4165--4174, 2014.

\bibitem{Kwon14}
Y.~Kwon, K.~Kang, and C.~Bae, ``Unsupervised learning for human activity
  recognition using smartphone sensors,'' \emph{Expert Systems with
  Applications}, vol.~41, no.~14, pp. 6067--6074, 2014.

\bibitem{Wang15}
W.~Wang, A.~X. Liu, M.~Shahzad, K.~Ling, and S.~Lu, ``Understanding and
  modeling of {WiFi} signal based human activity recognition,'' in
  \emph{Proceedings of the 21st Annual International Conference on Mobile
  Computing and Networking}.\hskip 1em plus 0.5em minus 0.4em\relax ACM, 2015,
  pp. 65--76.

\bibitem{Alvarez13}
A.~Alvarez-Alvarez, J.~M. Alonso, and G.~Trivino, ``Human activity recognition
  in indoor environments by means of fusing information extracted from
  intensity of wifi signal and accelerations,'' \emph{Information Sciences},
  vol. 233, pp. 162--182, 2013.

\bibitem{Brockmann06}
D.~Brockmann, L.~Hufnagel, and T.~Geisel, ``The scaling laws of human travel,''
  \emph{Nature}, vol. 439, no. 7075, pp. 462--465, 2006.

\bibitem{Gonzalez08}
M.~C. Gonz\'alez, C.~A. Hidalgo, and A.-L. Barab\'asi, ``Understanding
  individual human mobility patterns,'' \emph{Nature}, vol. 453, no. 7196, pp.
  779--782, 2008.

\bibitem{Song10}
C.~Song, Z.~Qu, N.~Blumm, and A.-L. Barabási, ``Limits of predictability in
  human mobility,'' \emph{Science}, vol. 327, no. 5968, pp. 1018--1021,
  February 2010.

\bibitem{Wang14}
R.~Wang, F.~Chen, Z.~Chen, T.~Li, G.~Harari, S.~Tignor, X.~Zhou, D.~Ben-Zeev,
  and A.~T. Campbell, ``Studentlife: assessing mental health, academic
  performance and behavioral trends of college students using smartphones,'' in
  \emph{Proceedings of the 2014 ACM International Joint Conference on Pervasive
  and Ubiquitous Computing}.\hskip 1em plus 0.5em minus 0.4em\relax ACM, 2014,
  pp. 3--14.

\bibitem{Weka}
{Machine Learning Group at the University of Waikato}, ``{Weka},''
  http://www.cs.waikato.ac.nz/ml/weka/, 2016.

\end{thebibliography}


\end{document}